\documentclass[twocolumn]{aastex62}

\usepackage{apjfonts}

\submitjournal{AJ}

\shorttitle{PNe in Globular Clusters}
\shortauthors{Bond et al.}

\newcommand{\Gaia}{{\it Gaia}}

\newcommand{\HST}{{\it HST}}

\newcommand{\kms}{{\>\rm km\>s^{-1}}}

\newcommand{\macd}{$\mu_\alpha \cos \delta$}
\newcommand{\md}{$\mu_\delta$}
\newcommand{\mv}{$m_{\rm F606W}$}

\newcommand{\mvi}{$m_{\rm F606W}-m_{\rm F814W}$}

\begin{document}

\title{Proper-Motion Membership Tests for Four Planetary Nebulae in Galactic Globular Clusters\footnote{Some of the data presented in this paper were obtained from the Mikulski Archive for Space Telescopes (MAST) at the Space Telescope Science Institute, operated by the Association of Universities for Research in Astronomy, Inc., under NASA contract NAS5-26555.
} 
}

\author[0000-0003-1377-7145]{Howard E. Bond}
\affil{Department of Astronomy \& Astrophysics, Pennsylvania State University, University Park, PA 16802, USA; heb11@psu.edu}
\affil{Space Telescope Science Institute, 
3700 San Martin Dr.,
Baltimore, MD 21218, USA}

\author[0000-0003-3858-637X]{Andrea Bellini}
\affil{Space Telescope Science Institute, 
3700 San Martin Dr.,
Baltimore, MD 21218, USA}

\author[0000-0001-6008-1955]{Kailash C. Sahu}
\affil{Space Telescope Science Institute, 
3700 San Martin Dr.,
Baltimore, MD 21218, USA}

\correspondingauthor{Howard E. Bond}
\email{heb11@psu.edu}

\begin{abstract}

Four planetary nebulae (PNe) are considered to be probable {or possible} members of Galactic globular clusters (GCs). These are Ps~1 = K648 in M15, GJJC~1 = IRAS\,18333$-$2357 in M22, JaFu~1 in Palomar~6, and JaFu~2 in NGC\,6441. In addition to lying close to the host GCs on the sky, these PNe have radial velocities that are consistent, within the errors and stellar velocity dispersions, with cluster membership. The remaining membership criterion is whether the proper motions (PMs) of the central stars are in agreement with those of the host clusters. We have carried out the PM test for all four PNe. Two of the central stars---those of Ps~1 and GJJC~1---have PMs listed in the recent \Gaia\/ Data Release~2 (DR2). {We updated the PM of the Ps~1 central star to a more precise value using archival {\it Hubble Space Telescope\/} (\HST) frames.} Both PMs are statistically consistent with cluster membership. For the other two PNe, we used archival \HST\/ images to derive the PMs of their nuclei. For JaFu~2, there are \HST\/ images at several epochs, and the measured PM of the nucleus is in excellent agreement with that of the host cluster. For JaFu~1 the available archival \HST\/ images are less optimal {and the results are less conclusive}; the measured PM for the central star is marginally consistent with cluster membership, {but additional astrometric observations are desirable for a more robust membership test.}

\end{abstract}

\keywords{planetary nebulae --- globular clusters}


\section{The Puzzle of Planetary Nebulae in Globular Clusters\label{sec:intro}}

Over nine decades ago, \citet{Pease1928} announced his discovery of a planetary nebula (PN) belonging to the globular cluster (GC) M15. The star K\"ustner~648 (K648) had attracted his attention because of its very blue color, and a follow-up spectrogram obtained at the Mount Wilson 100-inch telescope revealed an emission-line spectrum typical of a PN, superposed on the continuum of the blue star. After another six decades a second PN in a GC, this time belonging to M22, was discovered by \citet{Gillett1989} in the course of an investigation of infrared sources in the cluster. The infrared source is cataloged as IRAS\,18333$-$2357, and the PN is designated GJJC~1. In the 1990s a systematic search for PNe in Galactic GCs was conducted by \citet[][hereafter J97]{Jacoby1997}; they used ground-based CCD imaging with a narrow-band [\ion{O}{3}] 5007~\AA\ filter to observe 133 GCs. The J97 survey revealed two more PNe, lying close to the GCs Palomar~6 (Pal~6) and NGC~6441. These PNe are designated JaFu~1 and JaFu~2. A recent deep integral-field spectroscopic survey of 26 Galactic GCs \citep{Gottgens2019} did not reveal any further PNe. 

The presence of PNe in GCs is a challenge to our understanding of stellar evolution. In such old populations, stars leave the asymptotic giant branch (AGB) with masses of about $0.53\,M_\odot$ \citep[e.g.,][]{Alves2000, Kalirai2009, Cummings2018}. {The theoretical post-AGB evolutionary timescales of such low-mass remnants are so long, according to older studies \citep[e.g.,][]{Schoenberner1983}, that any nebular material ejected at the end of the AGB phase has ample time to disperse before the central star becomes hot enough to ionize it.  Thus, the single stars now evolving in GCs would not be expected to produce any visible ionized PNe. However, more recent theoretical work \citep[e.g.,][]{Miller2016} suggests that post-AGB evolution, even at low remnant masses, may be fast enough to explain these objects. The fact that there are a few PNe known in GCs implies that they may arise from binary stars---either blue stragglers that merged near the main sequence, producing a more massive star with a faster evolutionary timescale, or those that underwent common-envelope events which rapidly removed the AGB envelope and exposed a hot core that could photoionize the ejecta. See J97, \citet{Jacoby2013, Jacoby2017}, \citet{Otsuka2015}, \citet{Bond2015}, \citet{Boffin2019}, and references therein, for further discussion of binary scenarios for the origin of these objects. On the other hand, the existence of these objects may support the more recent theoretical studies of single-star evolution.}

These evolutionary considerations make it important to confirm that the PNe actually are members of the clusters, rather than chance superpositions. In addition to the PN lying angularly close to the GC, it is necessary to confirm that it has a radial velocity (RV) consistent with that of the cluster. Another test is that the PN has an interstellar extinction similar to that of the cluster (although this test can be complicated by internal dust in the PN)\null. {In the case of GCs, most of which have low metal contents, the PN would also be expected to have low abundances of heavy elements if it is a member.}


The remaining test is to confirm that the proper motions (PMs) of the central stars of the PNe are consistent with those of cluster members. This criterion has not as yet, to our knowledge, been applied to {three of the four PNe described above. \citet{Cudworth1990} measured the PM of the central star of GJJC\,1 relative to nearby cluster members using photographic material covering nearly a century, and concluded that it was consistent with membership within the uncertainties.} However, the availability of space-based astrometry now allows the PM requirement to be tested with high precision, and it is the purpose of the study reported here to carry out this analysis. 

Table~\ref{table:pnn} lists J2000 coordinates for the four PN nuclei, in the reference frame of the \Gaia\/ Data Release~2 \citep[DR2;][]{Gaia2018}. For three of them (Ps~1, GJJC~1, and JaFu~2) the coordinates of the central stars are taken directly from \Gaia\/ DR2. The central star of JaFu~1 is too faint to be contained in DR2, and we have instead determined coordinates in the DR2 astrometric frame using images obtained with the {\it Hubble Space Telescope\/} (\HST), as described below. The apparent magnitudes in the \Gaia\/ $G$ bandpass are listed for three of the stars, taken directly from DR2. For
JaFu~1 we estimated the $G$ magnitude approximately from a $V$-band (F555W) \HST\/ image that contained several brighter nearby stars with DR2 magnitudes.

For two of the central stars, \Gaia\/ DR2 already lists their PMs, and we discuss them in the next two sections. {In the case of K648, we have also used archival \HST\/ images to obtain an independent and more precise measurement of its PM\null.} The other two do not have PMs measured in \Gaia\/ DR2, so we have used archival \HST\/ frames to determine them, as described in the subsequent two sections.

\begin{deluxetable*}{lccccc}
\tablewidth{0 pt}
\tablecaption{Central Stars of Planetary Nebulae in Globular Clusters
\label{table:pnn}
}
\tablehead{
\colhead{Name} &
\colhead{PNG} &
\colhead{Cluster} &
\colhead{R.A. [J2000]} &
\colhead{Dec. [J2000]} &
\colhead{{\it G\/} [mag]} 
}
\startdata
Ps 1 = K648       & PN G065.0$-$27.3 & M15      & 21 29 59.397 &   +12 10 26.26 & 14.27 \\
GJJC 1 = IRAS 18333$-$2357 & PN G009.8$-$07.5 & M22      & 18 36 22.862 & $-$23 55 19.74 & 14.48 \\
JaFu 1            & PN G002.1+01.7   & Pal 6    & 17 43 57.243 & $-$26 11 53.75 & 20.1  \\
JaFu 2            & PN G353.5$-$05.0 & NGC 6441 & 17 50 10.923 & $-$37 03 27.58 & 15.62 \\
\enddata
\end{deluxetable*}

\def\mua{\mu_\alpha}
\def\mud{\mu_\delta}
\def\masyr{\rm mas\,yr^{-1}}

\section{P{\footnotesize s} 1 (K648) in M15\label{sec:ps1}}

A very high cluster membership probability is already well established for Ps~1. Its large negative RV, agreeing well with the RV of the cluster, was demonstrated in the discovery paper by Pease, and numerous subsequent studies have confirmed this. \citet{Joy1949} measured RVs of $-115$ and $-129\,\kms$ from two spectrograms of the PN, and \citet{Rauch2002} used two high-resolution ultraviolet spectra of photospheric lines of the central star obtained with the Goddard High-Resolution Spectrograph on \HST\/ to measure RVs of $-128$ and $-133\,\kms$. More recently, \citet{Otsuka2015} measured a mean RV of $-116.89\pm0.41\,\kms$ from 122 emission lines of the PN in a high-resolution echelle spectrogram. The mean RV of the cluster was found to be $-106.76\pm0.25\,\kms$ by \citet[][hereafter B19]{Baumgardt2019} from the average velocity of hundreds of red-giant members.\footnote{The B19 compilation, cited several times in this paper, is primarily based on unpublished spectra of large numbers of cluster members, obtained with the ESO Very Large Telescope (VLT) and the Keck Telescopes, and supplemented by published RVs, in particular from the VLT MUSE compilation of \citet{Kamann2018}. See \citet{Baumgardt2018} for more details, including data reduction.} {The heavy-element content of the nebula is very low \citep[e.g.,][and references therein]{Otsuka2015}, consistent with the low metallicity of the host cluster.}

\subsection{\Gaia\/ DR2 Proper Motion}

\Gaia\/ DR2 gives a PM for the central star of $(\mua\cos\delta,\mud) = (-0.72\pm 0.34, -2.74\pm0.34)\,\masyr$. Its absolute parallax from DR2 is $0.22\pm0.17$~mas, with a fractional uncertainty too high to be useful in testing cluster membership. Both the PM and parallax have relatively large uncertainties for a star this bright. (Note the considerably smaller errors in the next section for the M22 star, which has nearly the same apparent magnitude.) This may have resulted from a slightly non-stellar image of the central star, or from the relatively bright surrounding nebulosity.

\begin{figure}
\centering
\includegraphics[width=\columnwidth]{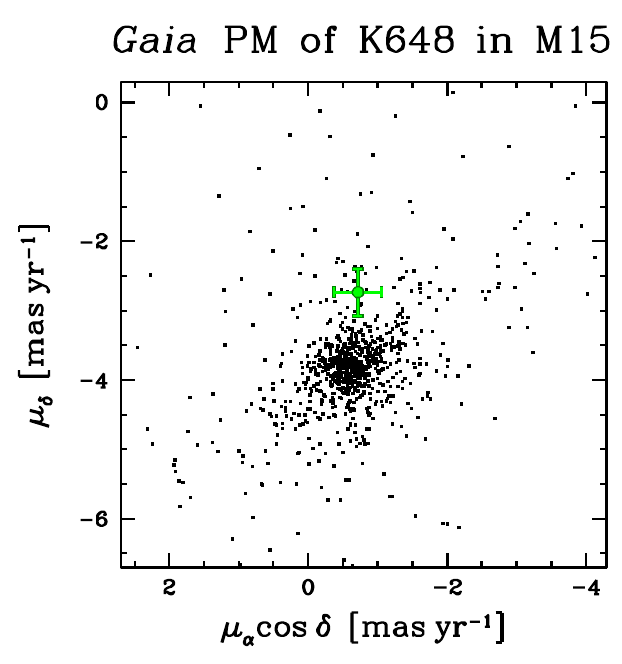}
\figcaption{Proper motions from \Gaia\/ DR2 for stars in the globular cluster M15 in the neighborhood of the planetary nebula Ps~1, selected as described in
the text. The green filled circle with error bars plots the proper motion of K648, the central star of Ps~1. 
\label{fig:Gaiam15}
}
\end{figure}

We selected a sample of stars in DR2 lying within
$120''$ of Ps~1, having magnitudes in the range $12<G<17$, and a parallax less than 4~mas. This sample contains a large percentage of cluster members, as shown by its \Gaia\/ color-magnitude diagram (CMD)\null. The PMs of this selection are plotted in Figure~\ref{fig:Gaiam15}. The PM of the central star, K648, is shown as a green point with error bars.

In this figure, the GC members are tightly clustered, but with a few outliers and/or field stars. The PM of the central star is statistically consistent with cluster membership, lying within about 2.5$\sigma$ of the mean cluster PM\null. 




\subsection{\HST\/ Proper Motion\label{sec:ps1HST}}


In order to determine the PM of K648 to higher precision than in \Gaia\/ DR2, we investigated archival \HST\/ images of M15, obtained from the Mikulski Archive for Space Telescopes (MAST)\null.\footnote{MAST is available at \url{http://archive.stsci.edu}} Extensive catalogs of PMs in M15 and other GCs determined from \HST\/ frames have been published by \citet[][hereafter B14]{2014ApJ...797..115B}.

Unfortunately, K648 is not present in the B14 catalog of M15 PMs, because its image was saturated in the reference frames they used. Therefore, we repeated the B14 PM reduction and analysis, this time including additional archival \HST\/ images in which K648 is unsaturated. Table~\ref{table:k648} lists the \HST\/ exposures of M15 we employed in this new analysis. As the table shows, these frames were obtained over a 19-year interval, from 1994 to 2013. The cameras used were the Wide Field Planetary Camera~2 (WFPC2), the High-Resolution and Wide-Field Channels of the Advanced Camera for Surveys (ACS/HRC and ACS/WFC), and the Ultraviolet-Visible channel of the Wide Field Camera~3 (WFC3/UVIS).



We made use of the \texttt{\_c0f} frames obtained from MAST for the WFPC2 observations, and the \texttt{\_flt} frames for ACS/HRC, ACS/WFC, and WFC3/UVIS\null. These frames are dark- and bias-subtracted, and have been flat-fielded, but no re-sampling has been applied; thus they preserve the full signal of the un-resampled pixel data for profile fitting. In addition, the \texttt{\_flt} frames for ACS/WFC and WFC3/UVIS have been pipeline-corrected to account for charge-transfer efficiency (CTE) losses, as described in detail in \citet{AndersonBedin2010}.\footnote{CTE losses are caused by charge traps in the CCD detectors that capture electrons and release them only after a delay into upstream pixels during the read-out process. As a result, stellar images have long ``tails.'' These shift the apparent positions of stars in the direction away from the readout register; the size of
this shift increases with distance from the readout register, fainter sources, and lower background levels. The effect also increases with time as the detector is exposed to cosmic radiation in the space environment.}

We derived image-tailored, empirical point-spread function (PSF) models by perturbing the library PSFs published by \citet[WFPC2]{AndersonKing2000}, \citet[ACS/WFC]{ak06}, \citet[ACS/HRC]{AndersonKing2006b}, and those made available online by J.~Anderson for the WFC3/UVIS camera.\footnote{\url{http://www.stsci.edu/~jayander/STDPSFs/}}
Only relatively bright and isolated stars that are present in each
image are used to perturb the PSF models (see \citealt{2017ApJ...842....6B}). When a library PSF is 
unavailable for a particular filter, we perturb the closest (in
wavelength) available model. We then used these PSF models with the
\texttt{FORTRAN} software package \texttt{hst1pass} (J.~Anderson, in
preparation) to measure initial stellar positions and fluxes in each
frame through a single pass of source finding, as described in \citet[][hereafter B18]{2018ApJ...853...86B}. We corrected the stellar positions in each catalog using state-of-the-art geometric distortion solutions as follows: for WFPC2: \citet{ak03}; for ACS: \citet{ak06, AndersonKing2006b}; for WFC3/UVIS: \citet{BelliniBedin2009} and \citet{Bellini2011}. 
We then defined a reference frame based on \Gaia\/ DR2 positions, oriented with north up, east on the
left.  We transformed single-exposure positions onto this reference
frame by means of six-parameter linear transformations. \Gaia\/
positions are shifted to the epoch of each data set to provide
smaller-residual transformations and to minimize mis-matching. Next,
we obtained our best estimates of stellar positions and fluxes using
the \texttt{KS2} package (see \citealt{2017ApJ...842....6B} for a
detailed description). \texttt{KS2} takes as input our image-tailored
PSF models, the \texttt{hst1pass}-based bright-star lists, and the
six-parameter transformations, and outputs deblended positions and
fluxes using all the exposures simultaneously, after passing through
several iterations of source finding.

We then obtained PMs by following the prescriptions given in B14 and B18. In a nutshell, for each source, we collect its $x$ and $y$ deblended positions as measured in each image, and transform them onto the reference frame. Transformed positions as a function of exposure epoch are then
iteratively fitted with a least-squares straight line, whose slope is a
direct measurement of the source's PM\null. Data rejection is a
critical part of the procedure; see B14 for details.

To mitigate the (generally small) uncorrected systematic residuals in
the measured stellar positions (e.g., lack of CTE correction for WFPC2 and ACS/HRC
exposures, lack of dithering in some datasets, uncorrected geometric-distortion residuals, and imperfect PSF models), source positions are locally transformed onto the reference frame using the
nearest 50 reference stars in each exposure. Our choice of reference
stars emphasizes cluster members, since their internal velocity
dispersion is much smaller than the field dispersion (i.e., smaller
transformation errors). As such, our PMs are necessarily
\textit{relative} to the bulk motion of the cluster. 
The final \HST\/ catalog contains over 130,000 sources.
The relative PM
of K648 we obtained from the \HST\/ data is $(\mu_\alpha\cos\delta,
\mu_\delta) = (+0.023\pm0.026, +0.332\pm 0.030)\,\masyr$. By assuming an absolute
cluster mean motion of $(-0.63\pm0.01, -3.80\pm0.01)\,\masyr$
(B19), we find an absolute PM of K648 of $(-0.607\pm0.028,
-3.468\pm0.032)\,\masyr$. This result improves the precision of the measurement by about an order of magnitude relative to \Gaia\/ DR2.



\begin{figure*}
\centering \includegraphics[width=\textwidth]{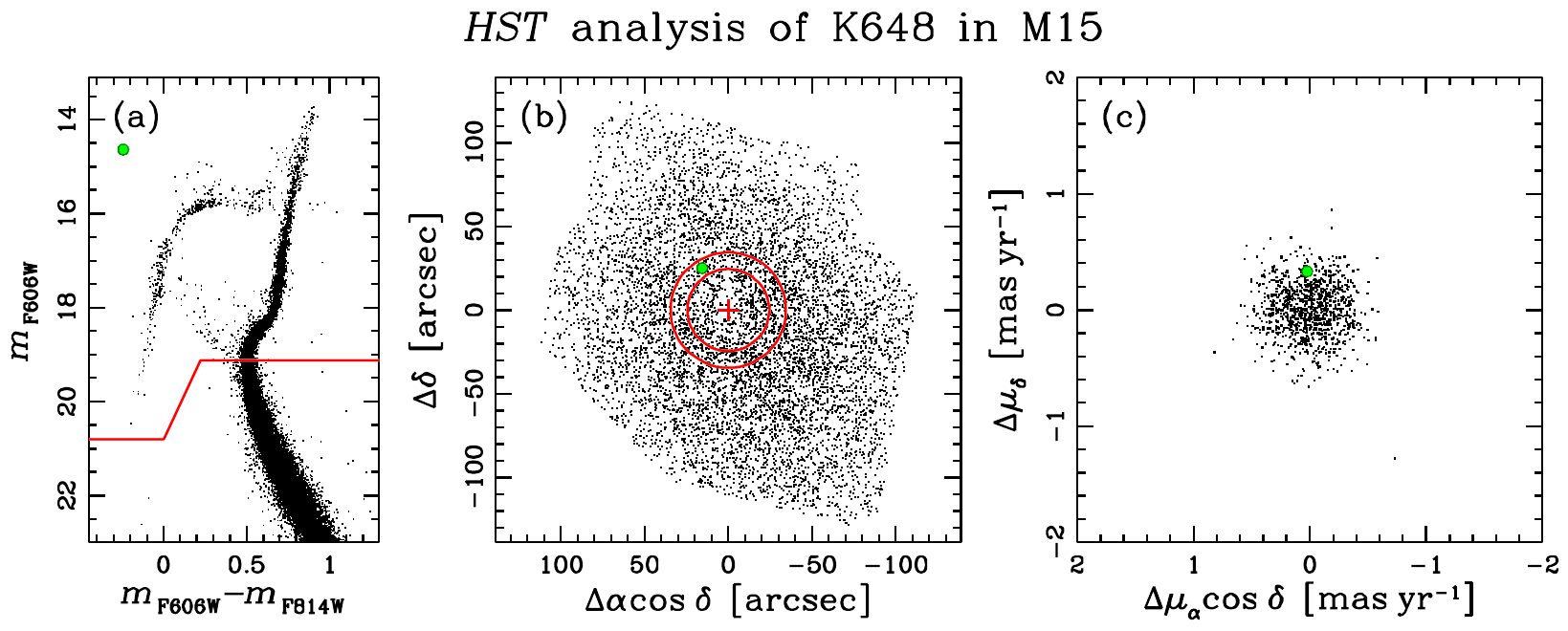}
\figcaption{
(a)~\mv\ versus \mvi\ color-magnitude diagram for well-measured sources in the \HST\/ proper-motion catalog of M15. The central star of Ps~1, K648, is marked by a green circle in all three panels. (b)~Map showing the field of view of the catalog relative to the center of the cluster, in units of arcsec. Only a 10\% randomly selected sample of well-measured stars is shown, for clarity. The cluster center is marked by the red cross. The red circles enclose a sample of stars whose distances from the cluster center are within $\pm\!5''$ of that of K648. (c)~Relative proper motions of evolved M15 stars (brighter than the red line in panel~a, and falling between the two red circles in panel~b). The uncertainties for K648 are smaller than the green plotting symbol, and are not shown. See text for details.\label{fig:HSTm15}}
\end{figure*}

Panel~(a) of Figure~\ref{fig:HSTm15} shows an \mv\ versus \mvi\ CMD of well-measured stars in the \textit{HST} field of M15. For this selection of stars, we isolated about 77,000 well-measured sources in the \HST\/ catalog, based on diagnostics provided by the reduction codes; these include the quality of the PSF fit and the reduced $\chi^2$ of the PM fits (see B14 and B18 for a detailed description of these diagnostics). The central star of Ps~1, K648, is plotted as a filled green circle in all three panels of Figure~\ref{fig:HSTm15}. It has measured magnitudes of $m_{\rm F606W}=14.631\pm0.031$ and $m_{\rm F814W}=14.872\pm0.027$. Panel~(b) in the figure shows a map of the field of view of well-measured stars in the PM catalog, with the center of M15 marked with a red cross. We selected a sample of stars with a distance from the cluster center within $\pm\!5''$ of the distance of K648, as shown by the two red circles in panel~(b). The reason for this restriction is that, due to the effects of hydrostatic equilibrium, stars closer to the cluster center have larger velocity dispersions than stars further out. 

Finally, panel~(c) shows the PM diagram for this distance-restricted sample. The stars in this panel have been further selected to exclude low-mass main-sequence stars by requiring them to lie above the red line in the CMD in panel~(a). This is because, due to the partial effects of energy equipartition (e.g., \citealt{tvdm13,bianchini16}), more-massive stars are kinematically cooler than the less-massive ones (see also Figure~21 of B14). The error bars on the PM of K648 are comparable in size to the plotting symbol, and are not shown.
As panel~(c) demonstrates, the PM of the central star is
kinematically consistent with it being a member of M15. 




\startlongtable
\begin{deluxetable*}{lcccc}
\tablewidth{0 pt}
\tablecaption{Archival \HST\/ Observations of K648
\label{table:k648} }
\tablehead{
\colhead{Date} &
\colhead{Program\tablenotemark{a}} &
\colhead{Camera} &
\colhead{Filter} &
\colhead{Exposure}
}
\startdata
1994 April 7      & 5324  & WFPC2     & F336W  & $2\times200$ s\\
                   &       &           & F439W  & $2\times30$ s\\
		  &	  &	      & F555W  & $4\times8$ s\\
\noalign{\smallskip}
1994 August 30    & 5687  & WFPC2     & F555W  & 100, 300 s\\
\noalign{\smallskip}
1994 October 26   & 5742  & WFPC2     & F336W  & $4\times200$, $7\times600$ s\\
\noalign{\smallskip}
1998 December 15  & 6751  & WFPC2     & F336W  & $9\times23$ s \\
1998 December 16  &       &           &        & 23 s\\
1998 December 17  &       &           &        & 23 s\\
1998 December 18  &       &           &        & 23 s\\
1998 December 20  &       &           &        & 23 s\\
1998 December 22  &       &           &        & 23 s\\
1998 December 15  &       &           & F439W  & $9\times20$ s \\
1998 December 16  &       &           &        & 20 s\\
1998 December 17  &       &           &        & 20 s\\
1998 December 18  &       &           &        & 20 s\\
1998 December 20  &       &           &        & 20 s\\
1998 December 22  &       &           &        & 20 s\\
1998 December 15  &       &           & F675W  & 400, $3\times500$ s \\
1998 December 15  &       &           & F814W  & $9\times14$ s \\
1998 December 16  &       &           &        & 14 s\\
1998 December 17  &       &           &        & 14 s\\
1998 December 18  &       &           &        & 14 s\\
1998 December 20  &       &           &        & 14 s\\
1998 December 22  &       &           &        & 14 s\\
\noalign{\smallskip}
1999 August 31    & 7469  & WFPC2     & F555W  & $12\times26$ s\\
1999 September 10 &       &           &        & $9\times260$ s\\
1999 August 31    &       &           & F785LP & $11\times26$ s\\
1999 September 10 &       &           &        & 120, $6\times260$ s\\
\noalign{\smallskip}
2004 December 6   & 10401 & ACS/HRC   & F435W  & $13\times125$ s\\
\noalign{\smallskip}
2006 May 2        & 10775 & ACS/WFC   & F606W  & 15, $4\times130$ s \\
                   &       &           & F814W  & 15, $4\times150$ s \\
\noalign{\smallskip}
2010 May 19       & 11233 & WFC3/UVIS & F390W  & $3\times827$ s\\
2010 May 20       &       &           &        & $3\times827$ s\\
\noalign{\smallskip}
2011 October 7    & 12605 & WFC3/UVIS & F336W  & $2\times350$ s\\
2011 October 16   &       &           &        & $2\times350$ s\\
2011 October 22   &       &           &        & $2\times350$ s\\
2011 October 7    &       &           & F438W  & $2\times65$ s\\
2011 October 16   &       &           &        & $2\times65$ s\\
2011 October 22   &       &           &        & $2\times65$ s\\
\noalign{\smallskip}
2012 September 17 & 12751 & WFC3/UVIS & F438W  & $4\times340$ s\\
                   &       &           & F606W  & $4\times47$ s\\
                   &       &           & F814W  & $4\times83$ s\\
\noalign{\smallskip}
2013 September 1  & 13295 & WFC3/UVIS & F555W  & $2\times10$ s\\
                   &       &           & F814W  & $2\times10$ s\\
\enddata
\tablenotetext{a}{PIs for these programs: B.~Yanny (5324); J.~Bahcall (5687); 
J.~Westphal (5742); H.~Bond (6751); W.~van~Altena (7469); R.~Chandar (10401); 
A.~Sarajedini (10775); G.~Piotto (11233 and 12605); C.~Heinke (12751); S.~Larsen 
(13295).}
\end{deluxetable*}

\null\bigbreak
\goodbreak
\bigbreak


\section{GJJC~1 (IRAS 18333$-$2357) in M22}

As in the case of Ps~1, the discovery paper for GJJC~1 = IRAS 18333$-$2357 \citep{Gillett1989} reported a large negative RV for the PN, close to that of the cluster, which again strongly supports cluster membership. They measured an RV of $-162\pm25\,\kms$ from the [\ion{O}{3}] emission lines, and $-157\pm15\,\kms$ from \ion{He}{2} absorption lines in the spectrum of the nucleus. B19 determined a mean cluster RV of $-147.76\pm0.30\,\kms$, based on measurements of bright member stars.

\begin{figure}
\centering
\includegraphics[width=\columnwidth]{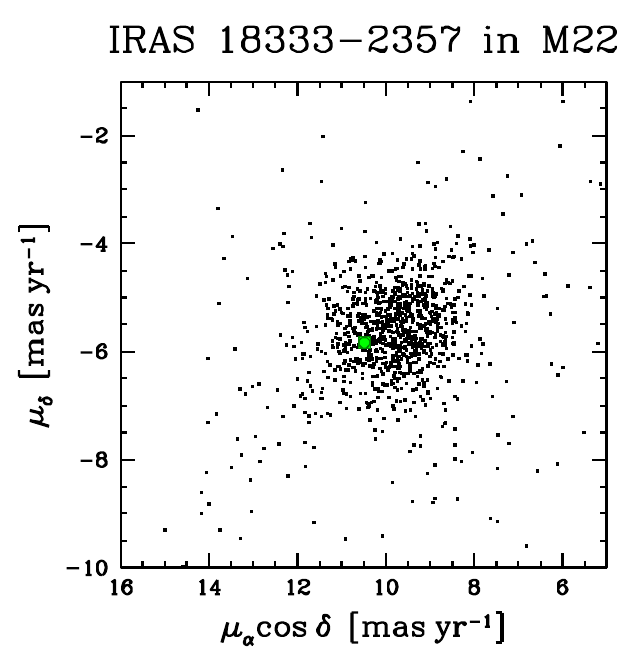}
\figcaption{Proper motions from \Gaia\/ DR2 for stars in the globular cluster M22 in the neighborhood of the planetary nebula GJJC~1, selected as
described in the text. The green point plots the proper motion of the central star of the nebula; its error bars, which are not plotted, are comparable in size to the plotting symbol.\label{fig:iras}}
\end{figure}

The \Gaia\/ DR2 PM for the nucleus of GJJC~1 is relatively large: $(\mua\cos\delta,\mud) = (+10.483\pm0.091, -5.835\pm0.076 )\,\masyr$. Its parallax is $0.2939\pm0.0643$~mas, consistent with cluster membership but not decisive. Similarly to M15, we selected a sample of stars in \Gaia\/ DR2 lying within
$120''$ of the PN, having magnitudes in the range $10<G<17$, and a parallax less
than 4~mas. The PMs of this selection are plotted in Figure~\ref{fig:iras}. The
PM of the central star is shown as a green filled circle, this time without error
bars since they are only slightly larger than the plotting symbol. 

The PM of the central star lies well within the distribution of cluster members, in accordance with it being a cluster member.\footnote{
The central star of GJJC\,1 is not contained in the \HST\/ PM compilation for M22 published by B14 because, as in the case of K648, it was saturated in the reference frames. Since the \Gaia\/ DR2 result is robust and conclusive, we did not attempt a reduction and analysis of the \HST\/ archival data.}
{The DR2 result is consistent with the conclusions of the photographic study by \citet{Cudworth1990}, mentioned in the introduction.}


\section{J{\footnotesize a}F{\footnotesize u} 1 in P{\footnotesize al} 6\label{sec:jafu1}}

J97 obtained a slit spectrogram of JaFu~1 that verified its PN nature. These authors measured a RV of $+176\pm15\kms$, which agrees extremely well with the mean
cluster RV of $+176.28\pm1.53\kms$ determined from about a dozen individual stars by B19. However, J97 pointed out that the PN lies at a relatively large separation of $230''$ from the cluster center (although this is still within the tidal radius).  Since
Pal~6 lies in a low-Galactic-latitude field, where the surface density of PNe is high, there is a possibility of a chance superposition. Moreover, J97 noted that the velocity dispersion in the surrounding Galactic bulge field is large enough that there is a small chance of a field PN having a
similar RV to that of the cluster. {By combining a non-membership probability of 3\% based on the agreement in RV with a 15\% probability of a chance spatial coincidence, J97 concluded that there was only a $\sim\!0.5\%$ likelihood that JaFu~1 is not a cluster member. They also determined a heavy-element content for the nebula that is only slightly sub-solar---making the composition indecisive for cluster membership, since the metallicities of the cluster and surrounding field are also approximately solar. The PM test of cluster membership for this object would be a useful addition.} 



As noted above (\S\ref{sec:intro}), the central star of JaFu~1 is not contained in \Gaia\/ DR2. We therefore measured its PM using archival \HST\/ images obtained from MAST\null. Unfortunately there are only two sets of single-orbit data that cover the location of JaFu~1; see Table~\ref{table:Jafu1} for details. The time baseline of these observations is only (and exactly) two years. 


Measuring a reliable PM with these particular \HST\/ data is {much more challenging than for K648}. The first-epoch exposures were taken with WFPC2, using broad-band ``$V$'' (F555W) and ``$I$'' (F814W) filters, and a narrow-band H$\alpha$ filter (F656N)\null. At the date of these observations, the WFPC2 had been on the spacecraft for more than 14~years. By this time the effects of CTE losses (see \S\ref{sec:ps1HST}) had become significant. No CTE correction algorithm has been developed for WFPC2 astrometry.
%
Moreover, the WFPC2 images were not dithered, making 
it harder to mitigate uncorrected geometric-distortion residuals.
Furthermore, empirical PSF library models (\citealt{AndersonKing2000}) are not available for the F656N filter. Finally, JaFu~1 was placed in the lower-resolution WF3 chip of WFPC2, with a relatively large pixel scale, about 2--4 times coarser than other \HST\/ imagers.


\begin{deluxetable}{lcccc}
\tablewidth{0 pt}
\tablecaption{Archival \HST\/ Observations of JaFu~1
\label{table:Jafu1} }
\tablehead{
\colhead{Date} &
\colhead{Program\tablenotemark{a}} &
\colhead{Camera} &
\colhead{Filter} &
\colhead{Exposure} 
}
\startdata
2008 March 14 & 11308 & WFPC2/WF3 & F555W & $2\times 160$ s \\
              &                        &          & F814W & $2\times 160$ s \\      
              &                        &          & F656N & $2\times 500$ s \\
\noalign{\smallskip}
2010 March 14 & 11558 & ACS/WFC  & F502N & $3\times 796$ s \\     
\enddata
\tablenotetext{a}{PI for both programs: O.~De Marco}
\end{deluxetable}

By contrast, the three second-epoch exposures were obtained with the higher-resolution ACS/WFC, but only in a single filter, the narrow-band [\ion{O}{3}] F502N bandpass. For ACS/WFC, as noted in \S\ref{sec:ps1HST}, a high-precision pixel-based CTE correction algorithm is available, and the ACS/WFC pixel scale is twice as fine as that of WFPC2/WF3. Unfortunately, however, the second-epoch exposures were again not dithered, and there are also no high-precision empirical PSF library models available for the ACS F502N filter (\citealt{ak06}).

To obtain the PM of the central star from the \HST\/ data, we followed the same procedures described in
\S\ref{sec:ps1} for the analysis of K648 in M15. However, unlike the case of M15, due to
a lack of appropriate reference stars in the field, local
transformations were defined by using all the available stars in the
images. Thus the resulting PMs are relative to the mean motion of
all of the stars in the field.

At the end of these reduction steps, we were able to measure PMs for 262 sources present in both the WFPC2/WF3 and ACS/WFC exposures. Stars in common between our PM catalog and \Gaia\/ DR2 were used to convert our relative PMs into absolute measurements. The resulting PM for the nucleus of JaFu~1 is $(\mu_\alpha\cos\delta, \mu_\delta) = (-6.32\pm1.63, -7.95\pm1.63)\,\masyr$.

\begin{figure*}
\centering
\includegraphics[width=\textwidth]{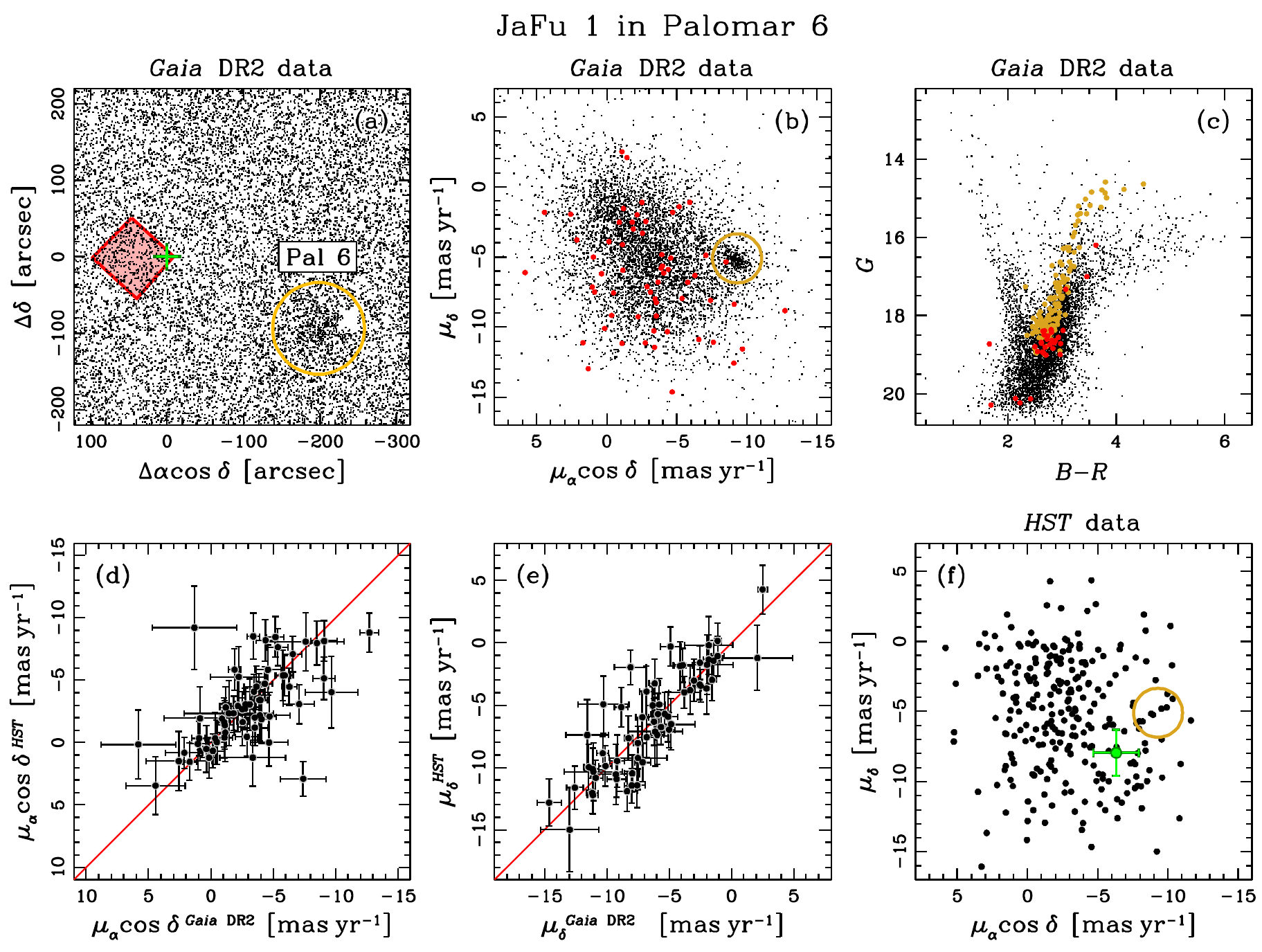}
\figcaption{(a) Positions of \textit{Gaia}~DR2 stars in the vicinity of Pal~6 (gold circle) and archival multi-epoch \textit{HST} exposures on JaFu~1 (red square). The position of JaFu~1 is marked with a green cross. Note a conspicuous dust cloud to the west-northwest side of the cluster. (b)~Vector-point PM diagram of \textit{Gaia} stars shown in panel (a). Red points are \textit{Gaia} stars within the \textit{HST} field of view. The gold circle highlights likely Pal~6 members according to their PMs. (c)~Color-magnitude for the \Gaia\/ stars. The gold points are likely cluster members, having positions and PMs within the gold circles in both panels~(a) and~(b). Red points are those lying within the \HST\/ field of view. (d) and (e) PMs in right ascension and declination for stars measured in the \HST\/ field and cataloged by \Gaia. These show good consistency within the errors. (f)~Vector-point diagram for the PMs we measured in the \HST\/ field. The PM of the nucleus of JaFu~1 is marked with a green circle with error bars. The gold circle for the cluster motion is the same as in panel~(b). See the text for details of these diagrams.\label{fig:jafu1}}
\end{figure*}

Panel (a) of Figure~\ref{fig:jafu1} plots the positions of \textit{Gaia} DR2 sources (black dots) in the vicinity of Pal~6, which is marked by a gold circle in the southwest corner of the frame. Positions are given in arcseconds relative to the location of the central star of JaFu~1, marked with a green cross. The red square corresponds to the region imaged by the first-epoch WFPC2/WF3 exposures. In panel~(b) we plot the catalog PMs of the \Gaia\/ sources. The location of a clump of bona-fide Pal~6 members is
highlighted by a gold circle of radius $1.75\,\masyr$. Red points in this panel mark \Gaia\/ stars that fall within the WFPC2/WF3 field of view. It should be noted that, based on PMs, there are few Pal~6 members lying in the WFPC2 field. Panel~(c) shows the \textit{Gaia}-based CMD of all the \Gaia\/
sources in panel~(a). The red points are the \Gaia\/ stars lying
within the WFPC2/WF3 field. The gold points are the likely Pal~6 cluster members, whose positions lie within the gold circle in panel~(a), and whose PMs fall within the gold circle in panel~(b). We see a clear cluster red-giant branch in this CMD\null. However, in the WFPC2 field, there are few if any Pal~6 red giants in the CMD.


To examine the impact of uncorrected systematic effects in our PMs, we compared our \HST\/ PM measurements with those in the \Gaia\/ DR2 catalog. The results are collected in panels~(d) and~(e), comparing the PMs in right ascension and declination, respectively. The red lines in panels~(d) and~(e) are the lines of equality, not fits to the data. It is reassuring that, overall, the points in both panels align along the diagonals, indicating that our measurements are consistent with those in the \Gaia\/ catalog. The scatter of the points along the red line is consistent with the error bars for the \md\ direction, but is somewhat larger along the \macd\ direction (which also happens to be the direction for which \textit{Gaia} PMs have larger errors). 

Finally, panel~(f) in Figure~\ref{fig:jafu1} shows the PM diagram for stars in the WFPC2 field, based on our astrometric analysis of the \textit{HST} frames. The gold circle is the same as that of panel~(b), and is used as a reference for Pal~6 membership. The green point with error bars is our PM measurement of the central star of JaFu~1. It is a PM outlier with respect to the Galactic bulge stars, but its PM lies about 2$\sigma$ away from the locus of Pal~6 stars.

We believe this effort has exhausted the utility of the existing \HST\/ images for assessing the cluster membership of JaFu~1 in Pal~6. A strong argument in favor of membership is the close agreement in RV, as mentioned at the beginning of this section. Moreover, our measured PM is {at least} marginally consistent with membership. Possibly arguing against membership is the fact that there are very few PM-based members of the cluster at the location of the PN in the outskirts of Pal~6, but there are numerous field stars of the Galactic bulge, as indicated by panel~(b) in Figure~\ref{fig:jafu1}. We conclude that more \HST\/ PM data from new observations would be desirable to more firmly assess the membership status of JaFu~1 in the cluster Pal~6.

%
%
%

%

\section{J{\footnotesize a}F{\footnotesize u} 2 in NGC 6441}

A slit spectrum of JaFu~2 obtained by J97 confirmed that it is a PN\null. Its RV was measured to be $+37\pm4.7\kms$. {The mean RV of the cluster is $+17.27\pm0.93\,\kms$, based on velocities of bright members (B19).
The velocity dispersion of evolved stars in NGC\,6441 at the radius of JaFu~2 is about 14--$15\,\kms$ \citep[][B19]{Watkins2015}. Thus, although the RV of JaFu~2 differs by $\sim$$20\,\kms$ from the cluster mean, it is compatible with cluster membership at the $\sim$$1.4\sigma$ level.}


J97 reported that the PN lies only $37''$ from the cluster center. They also noted that the inferred interstellar reddening of the PN is very similar to that of the cluster, {and that the nebula has a low heavy-element content based on their abundance analysis, which is more consistent with the host cluster than the bulge field}. They concluded that ``membership in NGC\,6441 is highly likely, but a proper-motion analysis is required to be certain.'' As noted above, \Gaia\/ DR2 gives a position for the PN's nucleus, but it does not list a PM.

Similarly to the situation for M15 and M22, there are extensive archival \HST\/ observations of NGC~6441, which are suitable for PM determinations. A detailed PM analysis of these data has been published by B14, based on \HST\/ data obtained at four epochs between 2003 and 2011, using the ACS/HRC, ACS/WFC, and WFC3/UVIS cameras.

However, once again the central star of the PN was not included in the
published catalog, this time because the existing data did not provide
at least two epochs of observations for it. We therefore determined
its PM based on the same frames used in the
B14 analysis, with the addition of newer WFC3/UVIS
frames obtained in 2014, which include JaFu~2. Table~\ref{table:Jafu2}
lists the \HST\/ exposures on NGC~6441 used in our new
analysis. Following the reduction procedures described in \S\ref{sec:ps1},
we obtained a relative PM of the JaFu~2 central star
of $(\mu_\alpha\cos\delta, \mu_\delta) = (+0.289\pm0.042,
+0.164\pm0.055)\,\masyr$. By adopting the mean cluster motion from the
B19 compilation of $(-2.51\pm0.03,
-5.32\pm0.03)\,\masyr$, we find an absolute central-star PM of
$(-2.22\pm0.05, -5.16\pm0.06)\,\masyr$.



\begin{figure*}
\centering
\includegraphics[width=\textwidth]{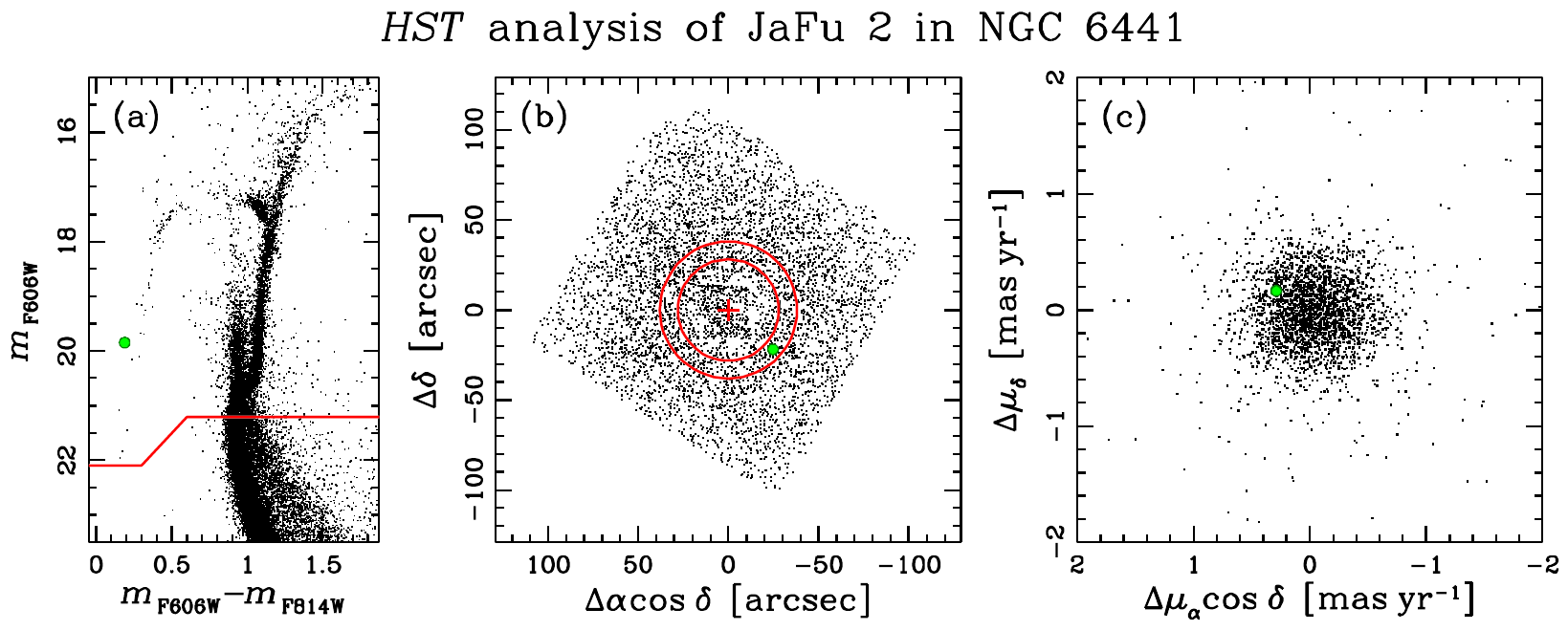}
\figcaption{
Similar to Figure~\ref{fig:HSTm15}, but for the cluster NGC\,6441 and the central star of JaFu~2. (a)~Color-magnitude diagram for the cluster. The central star is marked by a green circle in all three panels. (b)~Map of the positions of a 10\% random sample of cluster members; the cluster center is marked with a red cross, and the red circles enclose stars whose distances from the center are within $\pm\!5''$ of the distance of the central star. (c)~Proper motions of cluster stars inside the annulus in panel~b and lying above the red line in panel~a, and of the central star. Error bars for the central star are similar in size to the plotting symbol and are not shown. See text for details.\label{fig:jafu2}
}
\end{figure*}

Figure~\ref{fig:jafu2} is similar to Figure~\ref{fig:HSTm15}, but shows the data for
NGC~6441 and the central star of JaFu~2. Out of the over 140,000 entries in the \HST\/ PM catalog, the CMD in panel~(a) shows about 58,000 well-measured sources, selected as described in \S\ref{sec:ps1}. The central star (green circle in all panels) lies near the faint end of the blue horizontal branch. The measured magnitudes of the JaFu~2 nucleus are $m_{\rm F606W}=19.848\pm0.006$ and $m_{\rm F814W}=19.659\pm0.011$. 
A map of the field of view of the analyzed \HST\/ data is shown in panel~(b), with the cluster center marked by a red cross. The panel only shows a 10\% random sample of well-measured stars, for clarity. As for the case of K648, to better compare the kinematics of JaFu2 to that of cluster stars of similar properties, we further selected a subsample of evolved stars [brighter than the red line in panel~(a)], and lying within $\pm\!5^{\prime\prime}$ of the JaFu~2 distance of $33\farcs0$ from the cluster center [i.e., within the two red circles in panel~(b)]. (This value is an update of the $37''$ distance reported in the J97 discovery paper.) The vector-point diagram of this subsample is plotted in panel~(c). As in Figures~\ref{fig:HSTm15} and~\ref{fig:iras},
the error bars on the PM of JaFu~2 are comparable in size to the plotting symbol, and
are not shown. Our result is that the \HST\/ analysis reveals a PM of the JaFu~2 nucleus that is kinematically consistent with membership in NGC~6441.


%
%

\begin{deluxetable*}{lcccc}
\tablewidth{0 pt}
\tablecaption{Archival \HST\/ Observations of JaFu~2
\label{table:Jafu2} }
\tablehead{
\colhead{Date} &
\colhead{Program\tablenotemark{a}} &
\colhead{Camera} &
\colhead{Filter} &
\colhead{Exposure} 
}
\startdata
2003 September 1 & 9835  & ACS/HRC & F555W & $36\times240$ s \\
                 &                       &         & F814W & $5\times40$, \\
                 &                       &         &       & $2\times413$, \\
                 &                       &         &       & $10\times440$ s \\
\noalign{\smallskip}
2006 May 28      & 10775 & ACS/WFC & F606W & 45, $5\times340$ s \\  
                 &                       &         & F814W & 45, $5\times350$ s \\
\noalign{\smallskip}
2010 May 4       & 11739 & WFC3/UVIS& F390W & $3\times885$ s\\
2010 May 7       &   			 &  	    &	    & $3\times883$ s\\
2010 May 8       &   			 &  	    &	    & $3\times885$ s\\
2011 May 30      &                       &          &       & $3\times883$ s\\
\noalign{\smallskip}
2014 March 26    & 13297 & WFC3/UVIS& F336W & $2\times350$ s \\
                 &                       &          & F438W & 126, 128 s     \\
2014 June 15     &                       &          & F336W & 350 s          \\
                 &                       &          & F438W & 123 s          \\
2014 June 29     &                       &          & F336W & 350 s          \\
                 &                       &          & F438W & 129 s          \\
\enddata
\tablenotetext{a}{PIs for these programs: G.~Drukier (9835); A.~Sarajedini (10775); G.~Piotto (11739 and 13297).}
\end{deluxetable*}

\clearpage

\section{Summary}

Four PNe have been reported to be likely {or possible} members of Galactic GCs. The existence of PNe in GCs is difficult to understand in terms of single-star evolution, and may require an origin in binary interactions; {or it may support recent theoretical studies that report more rapid post-AGB evolution of single stars than found in older work.} It is important to verify that these PNe actually are members of the host clusters. We have used PMs of the central stars to test the cluster membership of these objects. 

Table~\ref{table:summary} summarizes our results. Column~2 gives the RVs measured for the PNe, quoted from the above sections; for GJJC\,1 we give the weighted mean of the two published values. Columns~3 and~4 list the absolute PMs of the central stars. Column~6 lists the mean RVs of the clusters, quoted from the recent compilation of B19.
Columns~7 and 8 give the mean PMs of the clusters, and column~9 lists the internal
central velocity dispersions of bright stars, all of these taken from the
B19 compilation. Cluster stars have fairly isotropic motions
in the central regions (e.g., \citealt{Watkins2015}), so that
line-of-sight and PM measurements of stars with similar kinematic masses
(as is generally the case for evolved stars brighter than the main-sequence turnoff) can be directly combined to provide more robust velocity-dispersion estimates.


Two of the central stars---in the clusters M15 and M22---have measured PMs in the recent \Gaia\/ DR2. {For the PN in M15, we updated the PM to a much more precise value, using archival \HST\/ frames.} The PMs of both central stars are seen to be statistically consistent with membership. 

For the other two PNe, for which \Gaia\/ DR2 does not give PMs, we determined the PMs of their central stars using archival \HST\/ images. For JaFu~1 in Pal~6, the \HST\/ material is less than ideal; {the RV agreement supports membership, but the other available information is equivocal.} It would be very desirable to obtain additional \HST\/ images to provide a better PM constraint. Our measurement of the PM of the nucleus of JaFu~2 in NGC~6441, based on excellent \HST\/ data, is fully consistent with cluster membership.

\begin{deluxetable*}{lcccccccc}
\tablewidth{0 pt}
\tabletypesize{\footnotesize}
\tablecaption{Summary of Proper-Motion and Radial-Velocity Data
\label{table:summary}
}
\tablehead{
\colhead{Central} &
\colhead{Radial} &
\colhead{$\mua\cos \delta$} &
\colhead{$\mud$} &
\colhead{Cluster} &
\colhead{Mean Radial} &
\colhead{Mean $\mua\cos \delta$} &
\colhead{Mean $\mud$} &
\colhead{$\sigma_\mu$} \\
\colhead{Star} &
\colhead{Velocity [$\kms$]} &
\colhead{[$\masyr$]} &
\colhead{[$\masyr$]} &
\colhead{} &
\colhead{Velocity [$\kms$]} &
\colhead{[$\masyr$]} &
\colhead{[$\masyr$]} &
\colhead{[$\masyr$]} 
}
\startdata
Ps 1 (K648)   & $-116.89\pm0.41$ & $-0.607\pm0.028$   & $-3.468\pm0.032$   & M15   & $-106.76\pm0.25$ & 
         $-0.63\pm0.01$ & $-3.80\pm0.01$   & $0.220\pm0.013$ \\
GJJC 1 & $-158\pm21$      & $+10.48\pm0.09$ & $-5.84\pm0.08$ & M22   & $-147.76\pm0.30$ & 
         $+9.82\pm0.01$    & $-5.54\pm0.01$ & $0.515\pm0.029$ \\
JaFu 1 & $+176\pm15$      & $-6.3\pm1.6$     & $-7.9\pm1.6$     & Pal\,6 & $+176.28\pm1.53$ & 
         $-9.17\pm0.06$   & $-5.26\pm0.05$   & $0.17\pm0.04$ \\
JaFu 2 & $+37\pm4.7$      & $-2.22\pm0.05$   & $-5.16\pm0.06$   & NGC\,6441 & $+17.27\pm0.93$ & 
         $-2.51\pm0.03$   & $-5.32\pm0.03$   & $0.264\pm 0.009$   \\
\enddata
\end{deluxetable*}

In summary, our study has strengthened the association of {three of} these four PNe with their host clusters, leaving intact their challenge to our understanding of low-mass stellar evolution. {The cluster membership of the fourth object, JaFu~1, remains somewhat uncertain, but is not ruled out.}

In a recent paper, \citet{Minniti2019} have identified four further cases of PNe lying close to the positions of GCs in the Galactic bulge. (A fifth candidate was ruled out on the basis of discordant RVs.) The GCs have been discovered in recent sky surveys. All four objects lie in extremely crowded star fields, making the probability of chance alignments relatively high. None of the objects have been imaged with \HST, and only one of them has an identified central star. It would be worthwhile to make efforts to identify the other objects' central stars, measure the RVs of the clusters and PNe, and to make PM membership studies similar to the one reported here.

\acknowledgments

Support for programs GO-13297 (A.B.) and GO-14794 (H.E.B.) was provided by NASA through grants from the Space Telescope Science Institute, which is operated by the Association of Universities for Research in Astronomy, Inc.

This work has made use of data from the European Space Agency (ESA) mission {\it Gaia\/} (\url{https://www.cosmos.esa.int/gaia}), processed by the {\it Gaia\/} Data Processing and Analysis Consortium (DPAC, \url{https://www.cosmos.esa.int/web/gaia/dpac/consortium}). Funding for the DPAC has been provided by national institutions, in particular the institutions 
participating in the {\it Gaia\/} Multilateral Agreement. 

\facilities{HST (WFPC2, ACS, WFC3), Gaia}


\end{document}